\documentclass{article}
\usepackage{ijcai16}

\usepackage{times}
\usepackage{graphicx}
\usepackage{amsmath}
\usepackage{amsfonts}
\usepackage{amssymb}
\usepackage{multirow}
\usepackage{subcaption}
\usepackage{url}
\usepackage{footmisc}

\title{Multimodal Emotion Recognition Using Multimodal Deep Learning}
\author{Wei Liu, Wei-Long Zheng, Bao-Liang Lu\thanks{Corresponding author} \\
Center for Brain-like Computing and Machine Intelligence\\
Department of Computer Science and Engineering \\
Key Laboratory of Shanghai Education Commission for \\
Intelligent Interaction and Cognition Engineering \\
Shanghai Jiao Tong University, Shanghai, China \\
\{liuwei-albert, weilong, bllu\}@sjtu.edu.cn}

\begin{document}

\maketitle

\begin{abstract}
  To enhance the performance of affective models and reduce the cost of acquiring 
  physiological signals for real-world applications, we adopt multimodal deep learning 
  approach to construct affective models from multiple physiological signals.
  For unimodal enhancement task, 
we indicate that the best recognition accuracy
of 82.11\% on SEED dataset is achieved with
shared representations generated by Deep AutoEncoder (DAE) model. 
  For multimodal facilitation tasks, we
  demonstrate that the Bimodal Deep AutoEncoder (BDAE) achieves the mean accuracies of 91.01\% and 83.25\% on SEED and DEAP datasets, respectively, which are  much superior to 
  the state-of-the-art approaches.
  For cross-modal learning task,
  our experimental results demonstrate that the mean accuracy of 66.34\% is achieved on SEED dataset through 
  shared representations generated by EEG-based DAE 
  as training samples and shared representations generated by eye-based
  DAE as testing sample, and vice versa.
\end{abstract}
\section{Introduction}
For human-machine interface (HMI), emotion recognition is becoming
more and more important.
Emotion recognition could be done through texts, pictures and physiological signals.
Bravo-Marquez {\it{et al.}} learned an expanded opinion (positive, neutral and negative) 
lexicon from emoticon annotated tweets~\cite{Bravo-Marquez:2015}.
Wang and Pal used constraint optimization framework to discover user' emotions
from social media content~\cite{wangdetecting}.\par 
Recently, many researchers studied emotion recognition from EEG. Liu {\it{et al.}} used
fractal dimension based algorithm to recognize and visualize emotions in real time~\cite{liu2010real}.
Murugappan {\it{et al.}} employed discrete wavelet transform to extract frequency
features from EEG signals and two classifiers are used to classify the features~\cite{murugappan2010classification}.
Duan {\it{et al.}} found that differential entropy features are more suited
for emotion recognition tasks~\cite{duan2013differential}. 
Zheng and Lu employed deep neural network
to classify EEG signals and examined critical bands and channels of EEG for 
emotion recognition~\cite{zheng2015investigating}.\par
Besides EEG signals, eye movement data can be used to find out what is attracting 
users' attention and observe users' unconscious behaviors. It is widely believed that
when people are in different emotions, the paradigm of eye movements and pupil
diameters will be different.
Nelson {\it{et al.}} studied the relationship between attentional bias to 
threat and anxiety by recording eye movement signals in different situations~\cite{nelson2015distinguishing}.
Bradley and Lang recorded eye movement signals to study the relationship
between memory, emotion and pupil diameters~\cite{bradley2015memory}.\par
To deal with information from different modalities, Yang {\it{et al.}} proposed
an auxiliary information regularized machine， which treats different modalities with 
different strategies~\cite{yang2015auxiliary}. Zhang {\it{et al.}} proposed a multimodal ranking aggregation 
framework for fusion of multiple visual tracking algorithms~\cite{zhang2015multi}.
In ~\cite{ngiam2011multimodal}, the authors built a single modal deep 
autoencoder and a bimodal deep autoencoder to generate shared representations of images and 
audios. Srivastava and Salakhutdinov extended the methods developed by 
 ~\cite{ngiam2011multimodal} to 
bimodal deep Boltzmann machines to handle multimodal deep learning problems~\cite{srivastava2014multimodal}.\par 
As for multimodal emotion recognition, 
Verma and Tiwary carried out emotion classification experiments with EEG singals
and peripheral physiological signals~\cite{verma2014multimodal}.
Lu {\it{et al.}} used two different fusion strategies for combining EEG and eye movement data: 
feature level fusion and decision level fusion~\cite{lu2015combining}. Their experimental results indicated that the best recognition accuracy was achieved by using fuzzy 
integral method in decision level fusion.
Vinola and Vimaladevi gave a detailed survey on human emotion recognition and listed
many other multimodal datasets and methods~\cite{vinola2015survey}.
\par
To our best knowledge, there is no research work reported in the literature dealing with emotion recognition from
multiple physiological signals
using multimodal deep learning algorithms. In this paper, we propose a novel multimodal emotion
recognition method using multimodal deep learning techniques.
In Section \ref{sec:model_arch}, we will introduce the unimodal deep autoencoder
and bimodal deep autoencoder. Section \ref{sec:setting} contains contents
about data pre-proessing, feature extraction and experiment settings. The experiment
results are described in Section \ref{sec:res}. Following 
discusses in Section \ref{sec:discussion}, conclusions and future work are represented in Section \ref{sec:conclusion}.
\section{Multimodal Deep Learning}
\label{sec:model_arch}
\subsection{Restricted Boltzmann Machine}
A restricted Boltzmann machine (RBM) is an undirected graph model, which has a visible
layer and a hidden layer. Connections exist only between visible layer and hidden layer
and there is no connection either in visible layer or in hidden layer. Assuming
visible variables $\mathbf{v}\in\{0,1\}^M$ and  hidden variables $\mathbf{h}\in\{0,1\}^N$,
we have the following energy function $E$:
\begin{equation}
  \label{equ:energy}
  E(\mathbf{v},\mathbf{h};\theta) = -\sum_{i=1}^M\sum_{j=1}^NW_{ij}v_ih_j-\sum_{i=1}^M
  b_iv_i - \sum_{j=1}^Na_jh_j
\end{equation}
where $\theta = \{\mathbf{a,b,W}\}$ are parameters, $W_{ij}$ is the symmetric weight
between visible unit $i$ and hidden unit $j$, $b_i,a_j$ are bias terms of visible
unit and hidden unit, respectively. With energy function, we can get the joint
distribution over the visible and hidden units:
\begin{align}
  \label{equ:joint}
  &p(\mathbf{v},\mathbf{h};\theta) = \frac{1}{\mathcal{Z}(\theta)}\exp(E(\mathbf{v},\mathbf{h};\theta))\\
  &\mathcal{Z}(\theta) = \sum_\mathbf{v}\sum_\mathbf{h}\exp(E(\mathbf{v},\mathbf{h};\theta))\nonumber
\end{align}
where $\mathcal{Z}(\theta)$ is the normalization constant. From Eqs. (\ref{equ:energy})
and (\ref{equ:joint}), we can derive the conditional distribution over hidden units $\mathbf{h}$ and
visible units $\mathbf{v}$ as follows:
\begin{align*}
  & p(\mathbf{h}|\mathbf{v};\theta) = \prod_{j=1}^Np(h_j|\mathbf{v})\\
  & p(\mathbf{v}|\mathbf{h};\theta) = \prod_{i=1}^Mp(v_i|\mathbf{h})
\end{align*}
with
\begin{align*}
  & p(h_j=1|\mathbf{v};\theta) = g\bigg(\sum_{i=1}^MW_{ij}v_i+a_j\bigg)\\
  & p(v_j=1|\mathbf{h};\theta) = g\bigg(\sum_{j=1}^NW_{ij}h_j+b_i\bigg)\\
  & g(x) = 1 / (1+\exp(-x))
\end{align*}
Given a set of visible variables $\{\mathbf{v}_n\}_{n=1}^N$, the derivative of log-likelihood
with respect to weight $\mathbf{W}$ can be calculated from Eq. (\ref{equ:joint}):
\[
 \frac{1}{N}\sum_{i=1}^N\frac{\partial \log P(\mathbf{v}_n;\theta)}{\partial W_{ij}}
 = \mathbb{E}_{P_{data}}[v_ih_j]-\mathbb{E}_{P_{model}}[v_ih_j]
\]
In this paper, we use Contrastive Divergence (CD) algorithm~\cite{hinton2002training} or
Persistent CD algorithm ~\cite{tieleman2008training} to train a RBM.
\subsection{Model construction}
\begin{figure*}[!ht]
    \centering
    \begin{subfigure}[]{0.4\textwidth}
        \includegraphics[width=\textwidth]{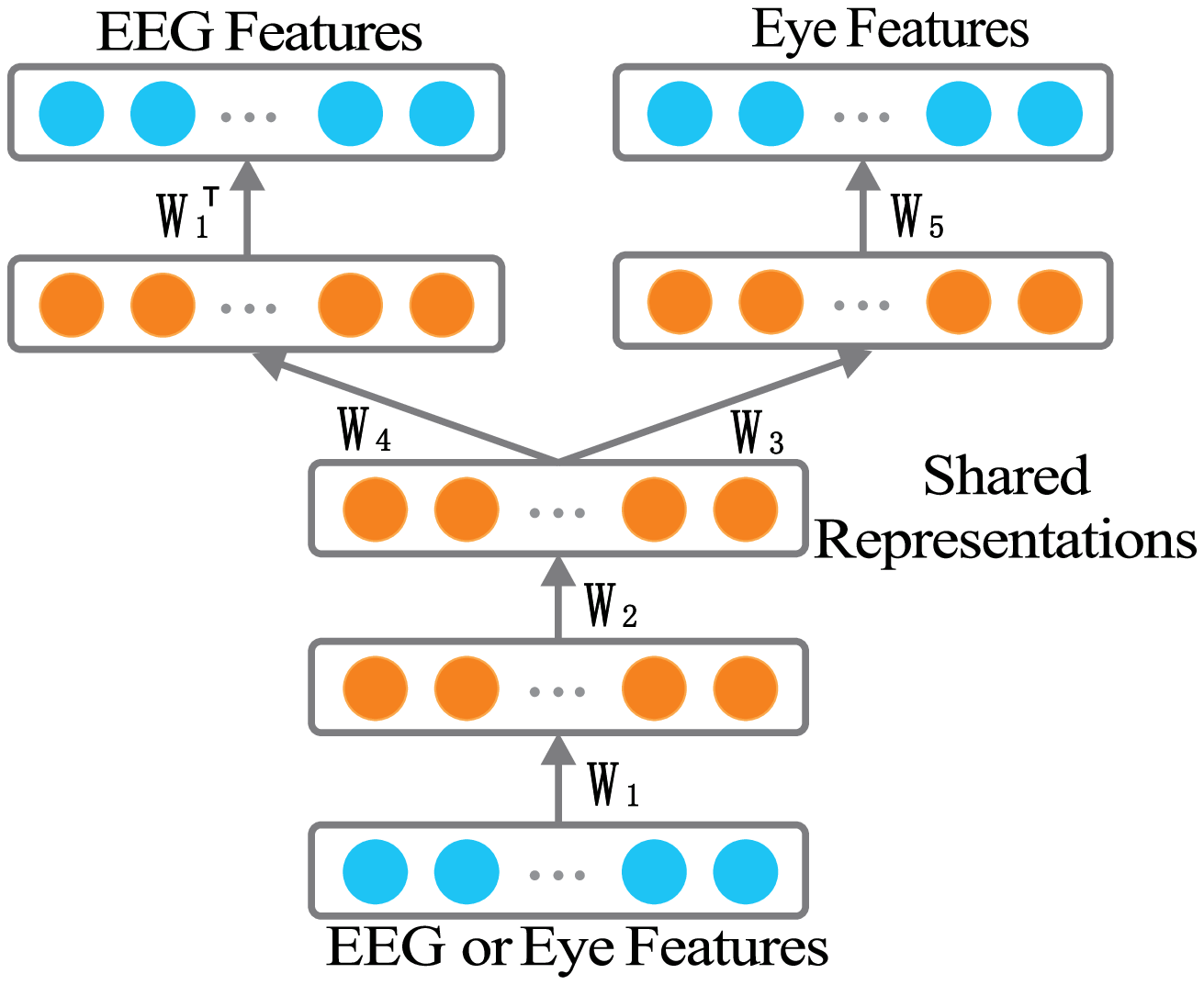}
        \caption{}
        \label{fig:dae}
    \end{subfigure}
    \hfil
    \begin{subfigure}[]{0.4\textwidth}
        \includegraphics[width=\textwidth]{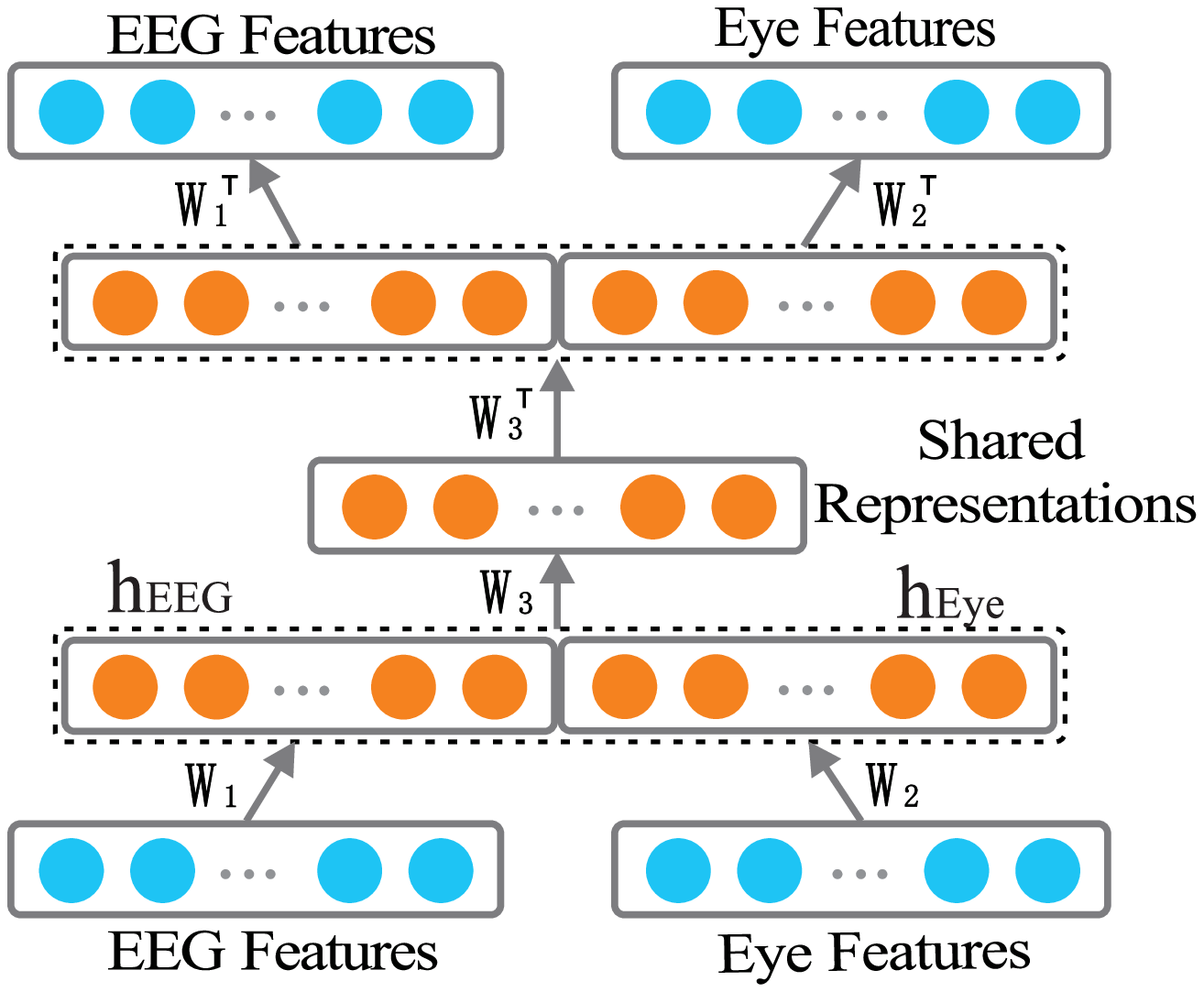}
        \caption{}
        \label{fig:bdae}
    \end{subfigure}
    \caption{Deep autoencoder models adopted in this paper. Figures \ref{fig:dae} and \ref{fig:bdae} depict the structure of unimodal DAE and the structure of BDAE, respectively. For DAE model, the inputs are EEG features or eye movment features.
    For BDAE model, the inputs are both EEG features and eye movement features. The middle layers
    in both networks are shared representations.}
    \label{fig:model}
\end{figure*}

To enhance emotion recognition accuracy by combining
EEG and eye movement data, we adopt a Bimodal Deep autoencoder (BDAE)
to extract shared representations of EEG and eye movement data. When only one modality
is available, the unimodal deep autoencoder (DAE) is applied to extract shared
representations. These two kinds of deep autoencoder models are depicted in Figure \ref{fig:model}.\par
\subsubsection{BDAE training}
To train BADE, we first trained two RBMs for EEG signals and eye movement data,
respectively, i.e., the  first two layers in Figure \ref{fig:bdae}.
After training respective RBMs, two hidden layers indicated by $\mathrm{h_{EEG}}$ and $\mathrm{h_{Eye}}$ were linked together
directly and we treated the joint hidden layer as the visible layer of an upper
RBM. When unfolding the stacked RBMs into a bimodal deep autoencoder, we kept the weights 
tied. From Figure \ref{fig:bdae}, we can see that $W_1,W_2,W_3$ and $W_1^T,W_2^T,W_3^T$
were tied weights.
At last, we used unsupervised back-propagation algorithm to finely tune the weights and bias.
\subsubsection{DAE training}
A similar method was used when training DAE.
Only one RBM was constructed for EEG features or eye tracking features, and the hidden layer
of the first RBM was treated as the visible layer of the upper RBM.
However, when unfolding the stacked RBMs,
we only kept the weights of first EEG (or eye) layer and top EEG (or eye) layer
tied. From Figure \ref{fig:dae}, we can see that $W_1$ and $W_1^T$ were tied weights
while other weights were not.
Other weights and bias could be trained with CD algorithm or
Persistent CD algorithm. Unsupervised back-propagation
was also needed to finely tune the parameters.
\par
There are three steps in total. The first step is to train the DAE network
or the BDAE network. It is worth noting that
both modality information is needed when training those autoencoders. We will call this
step feature selection. The second step is supervised training.
After training autoencoders,
we can use them to generate shared representations and these shared representations
can then be used to train a classifier. And the last step is a testing process, from which
the recognition results are produced.

\section{Experiment settings}
\label{sec:setting}
\subsection{Dataset}
Two public datasets, the SEED dataset\footnote{\url{http://bcmi.sjtu.edu.cn/~seed/index.html}} and the DEAP dataset\footnote{\url{http://www.eecs.qmul.ac.uk/mmv/datasets/deap/readme.html}}, were used in this paper.
The SEED dataset was first introduced in ~\cite{zheng2015investigating}.
This dataset contains EEG singals and eye movement signals from 15 subjects
during watching emotional movie clips.
The dataset contains 15 movie clips and each clip lasts about 4 minutes long. The
EEG signals are of 62 channels at a sampling rate of 1000 Hz and the eye movement
signals contain information about blink, saccade fixation and so on.
In order to compare our proposed method with ~\cite{lu2015combining},
we use the same data as in ~\cite{lu2015combining}, that is, 27 data files from 9 subjects.
For every data file, the data from the subjects watching the first 9 movie clips are used as
training samples and the rest are used as test samples.\par
The DEAP dataset was first introduced in
~\cite{koelstra2012deap}. The EEG signals and peripheral physiological signals
of 32 participants were recorded when they were watching
music videos. The dataset contains 32 channel EEG signals and 8 peripheral
physiological signals. The emotional music videos include 40 one-minute long
small clips and subjects were asked to do self-assessment by assigning values from
1 to 9 to five different status, namely, valence, arousal, dominance, liking and familiarity.
In order to compare the performance of our proposed method with previous results in ~\cite{rozgic2013robust} and ~\cite{li2015eeg},
we did not take familiarity into consideration. We divided the trials into two
different classes according to the assigned values. The threshold we chose is 5, and
the tasks can be treated as four binary classification problems, namely, high
or low valence, arousal, dominance and liking. Among all of the data, 90\% samples were used as
training data and the rest 10\% samples were used as test data.
\par

\subsection{Feature Extraction}
\subsubsection{SEED dataset}
Power Spectral Density (PSD) and Differential Entropy (DE) features were extracted
from EEG data. Both two kinds of features contain five frequency bands:
delta (1--4Hz), theta (4--8Hz), alpha (8--14Hz), beta (14--31), and gamma (31--50Hz).
As for eye movement data, we used the same features as in ~\cite{lu2015combining},
which were listed in Table \ref{tab:eye_features}. The extracted EEG features and
eye movement features were then scaled between 0 and 1 and the scaled features were
used as the visible units of BDAE or DAE network.
\begin{table}[!ht]
\renewcommand{\arraystretch}{1.0}
\centering
\begin{tabular}{|l|l|}
\hline
 Eye movements parameters & Extracted features \\
 \hline
 \multirow{4}{*}{Pupil diameter(X and Y)} & Mean,standard deviation, \\
                                          & DE in four bands\\
                                          & (0--0.2Hz,0.2--0.4Hz,\\
                                          & 0.4--0.6Hz,0.6--1Hz)\\
 \hline
 Disperson(X and Y) & Mean, standard deviation \\
 \hline
 Fixation duration (ms) & Mean, standard deviation \\
 \hline
 Blink duration (ms) & Mean, standard deviation \\
 \hline
 \multirow{3}{*}{Saccade} & Mean, standard deviation of \\
                          & saccade duration(ms) and \\
                          & saccade amplitude($^\circ$)\\
\hline
\multirow{9}{*}{Event statistics} & Blink frequency,\\
                                  & fixation frequency,\\
                                  & fixation duration maximum,\\
                                  & fixation dispersion total,\\
                                  & fixation dispersion maximum,\\
                                  & saccade frequency,\\
                                  & saccade duration average,\\
                                  & saccade amplitude average,\\
                                  & saccade latency average.\\
\hline
\end{tabular}%
\caption{The details of the extracted eye movement features.}
\label{tab:eye_features}
\end{table}
\par
\subsubsection{DEAP dataset}
Instead of extracting features manually, we 
used the downloaded preprocessed data directly as the inputs of BDAE network and
DAE network to generate shared representations of EEG signals and
peripheral physiological signals. First, the EEG signals and peripheral 
physiological signals were separated and then the signals were segmented into 63 seconds.
After segmentation, we combined different channel data of the same time period (one second), forming
the input signals of BDAE network. At last, BDAE network generates shared
representations.

\subsection{Classification}
The shared representations generated by BDAE network or DAE network
were used to train a classifier. In this paper,
linear SVM was used.
Inspired by ~\cite{guo2005crossmodal},  we performed
experiments on the following three kinds of emotion recognition tasks to examine 
the efficiency of our proposed method.
\begin{itemize}
  \item[(1)] For unimodal enhancement task, we built a unimodal DAE network
  for EEG features or eye movement features to reconstruct both modalities.
  The mid-layer shared representations were extracted to train a classifier. In this
  task, only SEED dataset was used.
  \item[(2)] In multimodal facilitation task, both modalities were needed. The shared
  representations generated by BDAE network were fed into linear SVM to train
  a classifier. Both SEED dataset and DEAP dataset were used.
  \item[(3)] For cross-modal learning task, we built two unimodal DAEs
  for EEG features and eye movement features. The mid-layer outputs
  were extracted as shared representations. Then we used extracted features of one modality 
  as training samples and extracted features of the other modality as testing samples. 
  In this task, only SEED dataset was used.
\end{itemize}
\par

\section{Results}
\label{sec:res}
\subsection{Unimodal enhancement}
In unimodal enhancement task, we used EEG signals to reconstruct information of two 
modalities. Once the DAE network was trained, we could use it as a feature selector to 
generate shared representations, even if only one modality information
is available. For eye movement data, the process was the same
as when only EEG signals were available.\par
\begin{figure}
  \centering
  \includegraphics[width=.4\textwidth]{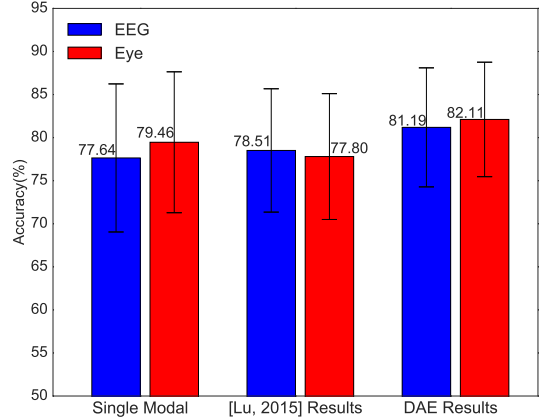}
  \caption{Summary of unimodal enhancement results. Three different models are compared. The
  left two bars are the results when only EEG features or Eye features are
  used. The middle bars show the results in \protect\cite{lu2015combining}, and the right
  two bars are the results of our DAE model. It is clear that our model
  performs best.}
  \label{fig:unimodal}
\end{figure}

Figure \ref{fig:unimodal} is the summary of all unimodal enhancement results.
We can see from Figure \ref{fig:unimodal} that the DAE
model performed best on both EEG features and eye movement features. 
\par
For EEG-based unimodal enhancement experiments, we constructed an affective model using EEG features 
of different frequency bands. The experimental results are shown in Table \ref{tab:eeg_only}.
After that an EEG-based DAE network was built to reconstruct both EEG and eye movement features
and the shared representations were used as new features to classify emotions.
The EEG-only DAE results are shown in Table \ref{tab:eeg_unimodal}.\par 
For eye-based unimodal enhancement experiments, the processes were the same. We 
carried out experiments using eye movement features. 
We linked all eye movement features listed in Table \ref{tab:eye_features} together to classify 
different emotions and the recognition accuracy of 79.64\% is achieved. Then, eye movement 
features were used to train the DAE network to reconstruct both EEG and eye movement 
features. Emotion recognition accuracies, as shown in Table \ref{tab:eye_unimodal}, were 
got with shared representations. \par  

\begin{table}[!ht]
 \centering
\renewcommand{\arraystretch}{1.0}
\setlength{\tabcolsep}{4pt}
\begin{tabular}{c|c|c|c|c|c|c|c}
\hline
 \multicolumn{2}{c|}{Feature} & $\delta$ & $\theta$ & $\alpha$ & $\beta$ & $\gamma$ & All \\
\hline
\hline
\multirow{2}{*}{PSD} & Ave. & 73.81 &	62.91 &	67.47 &	71.96 &	72.62 &	\textbf{77.54} \\
\cline{2-8}
   & Std. & 14.88 &	14.02 &	17.06 &	15.77 &	17.89 &	\textbf{12.62} \\
\hline
\hline
\multirow{2}{*}{DE} & Ave. & 70.97 &	67.98 &	71.91 &	75.47 &	\textbf{77.64} &	76.44 \\
\cline{2-8}
   & Std. & 15.35 &	15.64 &	16.32 &	15.57 &	17.19 &	\textbf{15.32} \\
\hline
\end{tabular}%
\caption{Recognition accuracy obtained by using EEG signal only. Here `All' represents the direct contatenation of all features from five frequency bands.}
\label{tab:eeg_only}
\end{table}
\begin{table}[!ht]
 \centering
 \renewcommand{\arraystretch}{1.0}
 \setlength{\tabcolsep}{4pt}
  \begin{tabular}{c|c|c|c|c|c|c|c}
  \hline
   \multicolumn{2}{c|}{Feature} & $\delta$ & $\theta$ & $\alpha$ & $\beta$ & $\gamma$ & All \\
  \hline
  \hline
  \multirow{2}{*}{PSD} & Ave. & \textbf{73.92} &	70.00 &	70.93 &	69.64 &	73.12 &	73.49 \\
  \cline{2-8}
     & Std. & 15.42 &	14.02 &	18.53 &	15.62 &	\textbf{11.12} &	16.51 \\
  \hline
  \hline
  \multirow{2}{*}{DE} & Ave. & 78.02 &	72.32 &	71.24 &	74.96 &	78.64 &	\textbf{81.19} \\
  \cline{2-8}
     & Std. & 12.55 &	15.52 &	14.58 &	13.91 &	\textbf{12.09} &	13.82 \\
  \hline
\end{tabular}
\caption{DAE--EEG features unimodal enhancement results. We used EEG data from different frequency bands to reconstruct both EEG and eye movement features. }
\label{tab:eeg_unimodal}
\end{table}%
\begin{table}[!ht]
 \centering
 \renewcommand{\arraystretch}{1.0}
 \setlength{\tabcolsep}{4pt} 
  \begin{tabular}{c|c|c|c|c|c|c|c}
  \hline
   \multicolumn{2}{c|}{Feature} & Re-$\delta$ & Re-$\theta$ & Re-$\alpha$ & Re-$\beta$ & Re-$\gamma$ & Re-All \\
  \hline
  \hline
  \multirow{2}{*}{PSD} & Ave. & 81.21 &	79.81 &	79.61 &	80.34 &	80.01 &	\textbf{82.11} \\
  \cline{2-8}
     & Std. & 13.30 &	13.61 &	13.35 &	14.24 &	\textbf{12.28} &	13.29 \\
  \hline
  \hline
  \multirow{2}{*}{DE} & Ave. & 81.19 &	81.00 &	\textbf{82.08} &	81.93 &	80.71 &	81.51 \\
  \cline{2-8}
     & Std. & \textbf{11.69} &	13.66 &	12.25 &	13.05 &	14.83 &	12.78 \\
  \hline
\end{tabular}
\caption{DAE--Eye movement features unimodal enhancement results. The prefix `Re' denotes reconstruct. Eye movement features were used to reconstruct EEG features and eye movement features.}
\label{tab:eye_unimodal}
\end{table}%
When only EEG features were used, 
we can see from Tables ~\ref{tab:eeg_only} and \ref{tab:eeg_unimodal}, the DAE network
increased the recognition rate from 77.64\% to 81.19\% and the standard deviation for
the best accuracy is 13.82, which is smaller than 17.19.
We also compared our results with ~\cite{lu2015combining}. In ~\cite{lu2015combining}, the 
best result achieved when only EEG signal used was 78.51\% and the standard 
deviation for its best accuracy was 14.32. It is clear that the DAE network is 
superior to the state-of-the-art approach.\par
When only eye movement data were available, the DAE network achieved the highest accuracy
of 82.11\% (in Table \ref{tab:eye_unimodal}) in comparison with the state-of-the-art 
approach ~\cite{lu2015combining} (77.80\%) and directly using eye movement features (79.46\%).

\subsection{Multimodal Facilitation}
We performed two kinds of different experiments to compare our BDAE network with other models.
\begin{itemize}
  \item[(1)] Only single modality is available.
  \item[(2)] When both modalities are available, the shared representations are obtained by
  linking the features directly.
\end{itemize}
\subsubsection{SEED results}
\begin{figure}
  \centering
  \includegraphics[width=.45\textwidth]{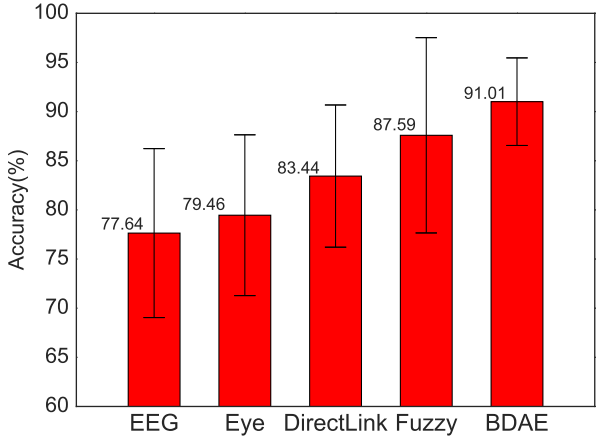}
  \caption{Multimodal facilitation results. The first two bars
   denote single modality. The rest bars denote multimodal with different fusion strategies.
   The fourth Fuzzy bar denotes the best result in \protect\cite{lu2015combining}.}
  \label{fig:multimodal}
\end{figure}

Figure \ref{fig:multimodal} shows the summary of
multimodal facilitation experiment results. We can see from Figure \ref{fig:multimodal}
that our BDAE model has the best performance (91.01\%). Besides, the standard deviation
of our BDAE model is also the smallest. This indicates that the BDAE model has a
good robustness.
Table \ref{tab:direct_link} shows the results when we linked the features extracted
from EEG signals and eye movement data directly.
The last column of Table \ref{tab:direct_link}
means that we linked both five frequency bands of EEG signals and eye movement
data features directly. \par
Compared with Table \ref{tab:eeg_only}, we can see that when linking different
modalities together, the emotion recognition accuracy increased in almost all frequency
bands, and the standard deviation becomes smaller.\par
\begin{table}[!ht]
\renewcommand{\arraystretch}{1.0}
\setlength{\tabcolsep}{3pt}
\begin{tabular}{c|c|c|c|c|c|c|c}
\hline
 \multicolumn{2}{c|}{Feature} & $\delta$+eye & $\theta$+eye & $\alpha$+eye & $\beta$+eye & $\gamma$+eye & All+eye \\
\hline
\hline
\multirow{2}{*}{PSD} & Ave. & 80.48 &	77.30 &	78.46 &	76.17 &	78.62 &	\textbf{80.88} \\
\cline{2-8}
   & Std. & 14.08 &	17.77 &	15.38 &	13.78 &	15.87 &	\textbf{13.47} \\
\hline
\hline
\multirow{2}{*}{DE} & Ave. & 83.33 &	81.79 &	79.14 &	82.98 &	\textbf{83.44} &	82.83 \\
\cline{2-8}
   & Std. & 13.44 &	13.47 &	\textbf{10.02} &	13.06 &	14.47 &	11.88 \\
\hline
\end{tabular}
\caption{Directly linking EEG and eye movement features.  The last column means we link all five frequency bands and eye movement features}
\label{tab:direct_link}
\end{table}
\begin{table}[!ht]
  \centering
  \renewcommand{\arraystretch}{1.0}
  \setlength{\tabcolsep}{3pt}
  \begin{tabular}{c|c|c|c|c|c|c|c}
  \hline
   \multicolumn{2}{c|}{Feature} & $\delta$+eye & $\theta$+eye & $\alpha$+eye & $\beta$+eye & $\gamma$+eye & All+eye \\
  \hline
  \hline
  \multirow{2}{*}{PSD} & Ave. & \textbf{85.12} &	83.89 &	83.18 &	83.23 &	82.92 &	85.10 \\
  \cline{2-8}
     & Std. & \textbf{11.09} &	13.13 &	12.68 &	13.65 &	13.59 &	11.82 \\
  \hline
  \hline
  \multirow{2}{*}{DE} & Ave. & 85.41 &	84.64 &	84.58 &	86.55 &	88.01 &	\textbf{91.01} \\
  \cline{2-8}
     & Std. & 14.03 & 11.03 &	12.78 &	10.48 &	10.25 &	\textbf{8.91} \\
  \hline
\end{tabular}
\caption{BDAE model results.}
\label{tab:BDAE}
\end{table}%
The experimental results using the BDAE model are shown in Table \ref{tab:BDAE}.
We examined the BDAE model three times and the recognition accuracies shown in Table \ref{tab:BDAE} were average.
We can see that the BDAE model achieved the best accuracy of 91.01\%,
which is higher than those of single modality and directly linking strategy.
And the standard deviation of the BDAE model is 8.91, which is the smallest among three different approaches.
\par
In ~\cite{lu2015combining}, the authors
employed fuzzy integral  method to fuse different modalities. The classification
accuracy is 87.59\% and the deviation is 19.87\%.
Compared with ~\cite{lu2015combining}, the BDAE model enhanced the performance of affective model significantly.\par
\subsubsection{DEAP results}
In previous papers, Rozgic {\it{et al.}} treated the EEG signals as a sequence of overlapping
segments and a novel non-parametric nearest neighbor model was employed to extract response-level feature from these segments~\cite{rozgic2013robust}.
Li {\it{et al.}} used Deep Belief Network (DBN) to automatically extract high-level
features from raw EEG signals~\cite{li2015eeg}.
\par 
The experimental results on the DEAP dataset are shown in Table \ref{tab:DEAP}.
We compared the BDAE results with results in ~\cite{li2015eeg} and
~\cite{rozgic2013robust}. As can be seen from Table \ref{tab:DEAP}, the BDAE
model improved recognition accuracies in all classification tasks.
\begin{table}[!ht]
  \renewcommand{\arraystretch}{1.0}
  \setlength{\tabcolsep}{3pt}
  \centering
  \begin{tabular}{l|c|c|c|c}
    \hline
          & Valence & Arousal & Dominance & liking \\ \hline
    ~\cite{rozgic2013robust} & 76.9 & 69.1 & 73.9 & 75.3 \\ \hline
    ~\cite{li2015eeg} & 58.4 & 64.3 & 65.8 & 66.9 \\ \hline
     \ \textbf{Our Method}  & \textbf{85.2}  & \textbf{80.5} & \textbf{84.9} & \textbf{82.4} \\ \hline
   \end{tabular}
  \caption{The experimental results on the DEAP dataset.(\%)}
  \label{tab:DEAP}
\end{table}
\par
From the experimental results on the SEED and DEAP datasets, we have demonstrated that the BDAE
network can be used to extract shared representations from different modalities
and the extracted features have better performance than other features.
\subsection{Cross-modal learning}
The key point of both DAE model and BDAE model is the shared representation.
In this section, the cross-modal experiments are
carried out to examine whether the shared representations can learn common information
between two different modalities.\par
In traditional machine learning framework, a classifier trained by EEG features
are usually considered to generate bad results when testing it on eye movement features.
\par
However, things are different when we use the DAE model. The DAE network is thought to
to able to learn something in common between different modalities. We
can test this by using shared representations generated by EEG features as training
samples and shared representations generated by eye features as testing samples, and
vice versa.\par
Both settings are examined, and the results are shown in Tables \ref{tab:eeg_train} and \ref{tab:eye_train}.
We first trained a classifier with shared representations generated from EEG fed
DAE network, and then tested the classifier with eye movement features generated shared
representations. The results are shown in Table \ref{tab:eeg_train}. As we can see from
Table \ref{tab:eeg_train}, both PSD features and DE features in all frequency
bands achieved accuracies more than 60\%, and the best performance is 66.23\%,
This is much higher than 33\% of random classification of three emotional states. \par
Then the other experiment setting was tested. We used shared representations generated
by eye movement features to train the classifier and the EEG-based shared representations
were used as testing samples. Table \ref{tab:eye_train} shows the results.
Similar to Table \ref{tab:eeg_train}, all accuracies are larger than 60\%, and
the best result is 66.45\%.
From Tables \ref{tab:eeg_train} and \ref{tab:eye_train}, we can see that the DAE models are able to learn
common features between EEG features and eye movement features. Though we do not
know what kind of shared representations they really are, we can take advantage
of this to improve emotion recognition accuracy.\par
\begin{table}[!ht]
\renewcommand{\arraystretch}{1.0}
\setlength{\tabcolsep}{4pt}
 \centering
\begin{tabular}{c|c|c|c|c|c|c|c}
\hline
 \multicolumn{2}{c|}{Feature} & $\delta$ & $\theta$ & $\alpha$ & $\beta$ & $\gamma$ & All \\
\hline
\hline
\multirow{2}{*}{PSD} & Ave. & 63.74 &	65.29 &	62.42 &	64.22 &	62.84 &	\textbf{66.23} \\
\cline{2-8}
   & Std. & 11.97 & \textbf{9.66} & 11.02 &	11.25 &	9.95 &	9.91 \\
\hline
\hline
\multirow{2}{*}{DE} & Ave. & 63.41 &	\textbf{66.08} &	61.52 &	64.42 &	63.33 &	65.82 \\
\cline{2-8}
   & Std. & 10.40 &	11.11 &	10.59 &	11.11 &	10.99 &	\textbf{8.32} \\
\hline
\end{tabular}
\caption{Cross-modal: EEG--training versus Eye--testing.}
\label{tab:eeg_train}
\end{table}%
\begin{table}[!ht]
  \renewcommand{\arraystretch}{1.0}
  \setlength{\tabcolsep}{4pt}
  \centering
  \begin{tabular}{c|c|c|c|c|c|c|c}
  \hline
   \multicolumn{2}{c|}{Feature} & Re-$\delta$ & Re-$\theta$ & Re-$\alpha$ & Re-$\beta$ & Re-$\gamma$ & Re-All \\
  \hline
  \hline
  \multirow{2}{*}{PSD} & Ave. & 62.07 &	65.69 &	\textbf{66.14} &	62.55 &	63.52 &	64.85 \\
  \cline{2-8}
     & Std. & 9.71 &	10.87 &	\textbf{8.57} &	9.83 &	11.08 &	10.79  \\
  \hline
  \hline
  \multirow{2}{*}{DE} & Ave. & 64.72 &	63.48 &	61.70 &	62.68 &	\textbf{66.45} &	63.57 \\
  \cline{2-8}
     & Std. & 11.83 &	10.16 &	9.77 &	9.87 &	\textbf{7.14} &	9.94 \\
  \hline
\end{tabular}
\caption{Cross-modal: Eye--training versus EEG--testing.}
\label{tab:eye_train}
\end{table}%

In the last, we analyzed the confusing matrices.
Table \ref{tab:matrix} shows the confusing matrices based on the experiment results
for each individual task. For convenient, we only listed the confusing matrices
on the SEED dataset. From Table \ref{tab:matrix}, we can see that negative emotions are
the hardest to recognize and positive emotions are easiest to recognize. This might
indicate that when people are happy or exciting, brain activities have some
common patterns while when people are sad, the patterns are not so obvious or the
patterns are changing with time.
\begin{table}[!ht]
  \renewcommand{\arraystretch}{1.0}
  \setlength{\tabcolsep}{7pt}
  \centering
  \begin{subtable}{.45\textwidth}
    \centering
    \begin{tabular}{|l|c|c|c|}
      \hline
                & Positive & Neutral & Negative \\
      \hline
       Positive & \textbf{99.03\%}  & 0.00\%  & 0.97\% \\
       \hline
       Neutral  & 3.7\%  & \textbf{90.26\%} & 6.03\% \\
       \hline
       Negative & 11.25\% & 3.57\% & \textbf{85.19\%} \\
       \hline
     \end{tabular}
    \caption{Multimodal Facilitation.}
  \end{subtable}
  \begin{subtable}{.45\textwidth}
    \centering
    \begin{tabular}{|l|c|c|c|}
      \hline
                & Positive & Neutral & Negative \\
      \hline
       Positive & \textbf{80.16\%}  & 9.73\%  & 10.11\% \\
       \hline
       Neutral  & 9.26\%  & \textbf{83.33\%} & 7.41\% \\
       \hline
       Negative & 9.24\% & 13.93\% & \textbf{76.83\%} \\
       \hline
     \end{tabular}
    \caption{Unimodal Enhancement}
  \end{subtable}
  \begin{subtable}{.45\textwidth}
    \centering
    \begin{tabular}{|l|c|c|c|}
      \hline
                & Positive & Neutral & Negative \\
      \hline
       Positive & \textbf{78.97\%}  & 10.61\%  & 10.42\% \\
       \hline
       Neutral  & 9.64\%  & \textbf{67.18\%} & 23.18\% \\
       \hline
       Negative & 25.14\% & 23.17\% & \textbf{51.68\%} \\
       \hline
     \end{tabular}
    \caption{Crossmodal Learning}
    \label{tab:crossmodal_matrix}
  \end{subtable}
  \caption{Confusing matrices of different tasks.}
  \label{tab:matrix}
\end{table}
\section{Discussion}
\label{sec:discussion}
All of three kinds of emotion recognition mentioned tasks above are important for HMI systems in practice. The multimodal
facilitation task allows us using different modalities so that HMI systems
could have a better recognition accuracy. Besides, the experiments results have indicated
that when both modalities were used, the standard deviation became smaller than
before. This phenomenon indicates that our system becomes more reliable.
Unimodal enhancement results have shown that if we train the DAE network with two modalities,
it is feasible to use only one modal in practice. Inspired by this results,
EEG signals might be not needed in practice and only some easily-collected signals
are used.
In the last, cross-modal learning tries to find out the common features
between EEG signals and eye movement data. The experiment results have demonstrated 
 that our shared representations do extract some common features between
EEG features and eye movement features.
\section{Conclusions and Future Work}
\label{sec:conclusion}
This paper has shown that by fusing EEG features and other features with bimodal
deep autoencoders (BDAE), the shared representations are good features to discriminate
different emotions. For the SEED dataset, compared with other feature merging strategies, 
the BDAE model
is better than others with the best accuracy of 89.94\%. In order to avoid intricacies
during acquiring EEG signals, we have adopted unimodal deep autoencoder model (DAE) to
extract shared representations even there was only one modality available. The experimental results on the unimodal
enhancement task have shown that the DAE model (82.11\%) performs better than
using single modality directly (78.51\% in ~\cite{lu2015combining}). In addition, the experimental results on the cross-modal learning
task demonstrated that the shared representations
contain higher level common features between EEG signals and eye movement features. The affective models with the shared representation performed much better than random classification (33.33\%) and achieved the
best accuracy of 66.45\%. \par 
As future work, we will focus on the following issues that 
we have not covered in this paper.
First, we will explore the relationship between unimodal features and shared
representations, so that we may find a clear explanation of confusing matrices.
Second, we want to go deeper with the performance of the DAE and BDAE networks when parameters
change.
Besides, more experiments are needed in order to study the stability of the DAE and BDAE
networks.
\section*{Acknowledgment}
This work was supported in part by the grants from the National Natural Science
Foundation of China (Grant No.61272248) and the National Basic Research Program of
China (Grant No.2013CB329401).

\bibliographystyle{named}
\bibliography{paper}

\begin{thebibliography}{}

\bibitem[\protect\citeauthoryear{Bradley and Lang}{2015}]{bradley2015memory}
Margaret~M Bradley and Peter~J Lang.
\newblock Memory, emotion, and pupil diameter: Repetition of natural scenes.
\newblock {\em Psychophysiology}, 52(9):1186--93, 2015.

\bibitem[\protect\citeauthoryear{Bravo-Marquez \bgroup \em et al.\egroup
  }{2015}]{Bravo-Marquez:2015}
Felipe Bravo-Marquez, Eibe Frank, and Bernhard Pfahringer.
\newblock Positive, negative, or neutral: Learning an expanded opinion lexicon
  from emoticon-annotated tweets.
\newblock In {\em IJCAI'15}, pages 1229--1235. AAAI Press, 2015.

\bibitem[\protect\citeauthoryear{Duan \bgroup \em et al.\egroup
  }{2013}]{duan2013differential}
Ruo-Nan Duan, Jia-Yi Zhu, and Bao-Liang Lu.
\newblock Differential entropy feature for eeg-based emotion classification.
\newblock In {\em 2013 6th International IEEE/EMBS Conference on Neural
  Engineering}, pages 81--84. IEEE, 2013.

\bibitem[\protect\citeauthoryear{Guo and Guo}{2005}]{guo2005crossmodal}
Jianzeng Guo and Aike Guo.
\newblock Crossmodal interactions between olfactory and visual learning in
  drosophila.
\newblock {\em Science}, 309(5732):307--310, 2005.

\bibitem[\protect\citeauthoryear{Hinton}{2002}]{hinton2002training}
Geoffrey~E Hinton.
\newblock Training products of experts by minimizing contrastive divergence.
\newblock {\em Neural Computation}, 14(8):1771--1800, 2002.

\bibitem[\protect\citeauthoryear{Koelstra \bgroup \em et al.\egroup
  }{2012}]{koelstra2012deap}
Sander Koelstra, Christian M{\"u}hl, Mohammad Soleymani, Jong-Seok Lee, Ashkan
  Yazdani, Touradj Ebrahimi, Thierry Pun, Anton Nijholt, and Ioannis Patras.
\newblock Deap: A database for emotion analysis using physiological signals.
\newblock {\em IEEE Transactions on Affective Computing}, 3(1):18--31, 2012.

\bibitem[\protect\citeauthoryear{Li \bgroup \em et al.\egroup
  }{2015}]{li2015eeg}
Xiang Li, Peng Zhang, Dawei Song, Guangliang Yu, Yuexian Hou, and Bin Hu.
\newblock {EEG} based emotion identification using unsupervised deep feature
  learning.
\newblock In {\em SIGIR2015 Workshop on Neuro-Physiological Methods in IR
  Research}, August 2015.

\bibitem[\protect\citeauthoryear{Liu \bgroup \em et al.\egroup
  }{2010}]{liu2010real}
Yisi Liu, Olga Sourina, and Minh~Khoa Nguyen.
\newblock Real-time {EEG}-based human emotion recognition and visualization.
\newblock In {\em 2010 International Conference on Cyberworlds}, pages
  262--269. IEEE, 2010.

\bibitem[\protect\citeauthoryear{Lu \bgroup \em et al.\egroup
  }{2015}]{lu2015combining}
Yifei Lu, Wei{-}Long Zheng, Binbin Li, and Bao{-}Liang Lu.
\newblock Combining eye movements and {EEG} to enhance emotion recognition.
\newblock In {\em IJCAI'15}, pages 1170--1176, 2015.

\bibitem[\protect\citeauthoryear{Murugappan \bgroup \em et al.\egroup
  }{2010}]{murugappan2010classification}
Murugappan Murugappan, Nagarajan Ramachandran, Yaacob Sazali, et~al.
\newblock Classification of human emotion from {EEG} using discrete wavelet
  transform.
\newblock {\em Journal of Biomedical Science and Engineering}, 3(04):390--396,
  2010.

\bibitem[\protect\citeauthoryear{Nelson \bgroup \em et al.\egroup
  }{2015}]{nelson2015distinguishing}
Andrea~L Nelson, Christine Purdon, Leanne Quigley, Jonathan Carriere, and
  Daniel Smilek.
\newblock Distinguishing the roles of trait and state anxiety on the nature of
  anxiety-related attentional biases to threat using a free viewing eye
  movement paradigm.
\newblock {\em Cognition and Emotion}, 29(3):504--526, 2015.

\bibitem[\protect\citeauthoryear{Ngiam \bgroup \em et al.\egroup
  }{2011}]{ngiam2011multimodal}
Jiquan Ngiam, Aditya Khosla, Mingyu Kim, Juhan Nam, Honglak Lee, and Andrew~Y
  Ng.
\newblock Multimodal deep learning.
\newblock In {\em ICML'11}, pages 689--696, 2011.

\bibitem[\protect\citeauthoryear{Rozgic \bgroup \em et al.\egroup
  }{2013}]{rozgic2013robust}
Viktor Rozgic, Shiv~N Vitaladevuni, and Ranga Prasad.
\newblock Robust {EEG} emotion classification using segment level decision
  fusion.
\newblock In {\em 2013 IEEE International Conference on Acoustics, Speech and
  Signal Processing}, pages 1286--1290. IEEE, 2013.

\bibitem[\protect\citeauthoryear{Srivastava and
  Salakhutdinov}{2014}]{srivastava2014multimodal}
Nitish Srivastava and Ruslan Salakhutdinov.
\newblock Multimodal learning with deep boltzmann machines.
\newblock {\em The Journal of Machine Learning Research}, 15(1):2949--2980,
  2014.

\bibitem[\protect\citeauthoryear{Tieleman}{2008}]{tieleman2008training}
Tijmen Tieleman.
\newblock Training restricted boltzmann machines using approximations to the
  likelihood gradient.
\newblock In {\em ICML'08}, pages 1064--1071. ACM, 2008.

\bibitem[\protect\citeauthoryear{Verma and Tiwary}{2014}]{verma2014multimodal}
Gyanendra~K Verma and Uma~Shanker Tiwary.
\newblock Multimodal fusion framework: A multiresolution approach for emotion
  classification and recognition from physiological signals.
\newblock {\em NeuroImage}, 102:162--172, 2014.

\bibitem[\protect\citeauthoryear{Vinola and
  Vimaladevi}{2015}]{vinola2015survey}
C~Vinola and K~Vimaladevi.
\newblock A survey on human emotion recognition approaches, databases and
  applications.
\newblock {\em ELCVIA: electronic letters on computer vision and image
  analysis}, pages 24--44, 2015.

\bibitem[\protect\citeauthoryear{Wang and Pal}{2015}]{wangdetecting}
Yichen Wang and Aditya Pal.
\newblock Detecting emotions in social media: A constrained optimization
  approach.
\newblock In {\em IJCAI'15}, pages 996--1002. AAAI Press, 2015.

\bibitem[\protect\citeauthoryear{Yang \bgroup \em et al.\egroup
  }{2015}]{yang2015auxiliary}
Yang Yang, Han-Jia Ye, De-Chuan Zhan, and Yuan Jiang.
\newblock Auxiliary information regularized machine for multiple modality
  feature learning.
\newblock In {\em IJCAI'15}, pages 1033--1039. AAAI Press, 2015.

\bibitem[\protect\citeauthoryear{Zhang \bgroup \em et al.\egroup
  }{2015}]{zhang2015multi}
Xiaoqin Zhang, Wei Li, Mingyu Fan, Di~Wang, and Xiuzi Ye.
\newblock Multi-modality tracker aggregation: from generative to
  discriminative.
\newblock In {\em IJCAI'15}, pages 1937--1943. AAAI Press, 2015.

\bibitem[\protect\citeauthoryear{Zheng and Lu}{2015}]{zheng2015investigating}
Wei{-}Long Zheng and Bao{-}Liang Lu.
\newblock Investigating critical frequency bands and channels for eeg-based
  emotion recognition with deep neural networks.
\newblock {\em {IEEE} Transactions on Autonomous Mental Development},
  7(3):162--175, 2015.

\end{thebibliography}

\end{document}